# Magnetic field-controlled lattice thermal conductivity in $MnBi_2Te_4$

Dung D. Vu[1,†], Ryan A. Nelson[4,‡], Brandi L. Wooten[3,§], Joseph Barker[5,∥], Joshua E. Goldberger[4,¶], Joseph. P. Heremans[1,2,3,#,*]

[1] Department of Mechanical and Aerospace Engineering, The Ohio State University, Columbus, OH 43210, USA.
[2] Department of Physics, The Ohio State University, Columbus, OH 43210, USA.
[3] Department of Materials Science and Engineering, The Ohio State University, Columbus, OH 43210, USA.
[4] Department of Chemistry and Biochemistry, The Ohio State University, Columbus, OH 43210, USA.
[4] Department of Chemistry and Biochemistry, The Ohio State University, Columbus, OH 43210, USA.
[5] School of Physics and Astronomy, University of Leeds, United Kingdom.
* Corresponding author (Email: heremans.1@osu.edu)
[†] ORCID 0000-0001-9085-0436
[‡] ORCID 0000-0003-0763-1094
[§] ORCID 0000-0003-4342-5976
[∥] ORCID 0000-0003-4843-5516
[¶] ORCID 0000-0003-4284-604X
[#] ORCID 0000-0003-3996-2744

**Abstract**

Phonon properties and the lattice thermal conductivity are not generally understood to be sensitive to magnetic fields. Using an applied field to change $MnBi_2Te_4$ between antiferromagnetic (AFM), canted (CAFM) and ferromagnetic (FM) phases we discovered a new way to control the lattice thermal conductivity, generating both a positive and a negative magnetic field dependence. We report the field dependence of the thermal conductivity, $\kappa$, in the in-plane direction under an applied magnetic field along the cross-plane direction in $MnBi_2Te_4$ from 2K to 30K. $\kappa$ decreases with field in the AFM phase, saturates in the CAFM phase, and increases with field in the FM phase. We explain this in terms of the field-induced changes of the magnon gap which modifies which magnon-phonon scattering processes are allowed by energy conservation. We also report magneto-Seebeck coefficient, Nernst coefficient and thermal Hall data measured in the same configuration. This finding may open a way to design magnetically controlled heat switches.

**Introduction**

The thermal conductivity of a solid has contributions from multiple heat carriers such as phonons, electrons and magnons. Magnetic fields affect these heat carriers, and hence the thermal conductivity, in different ways. Electrons carry both charge and spin, magnons carry spin, while phonons are usually considered both charge neutral and spinless. The thermal conductivity of the electrons can be modulated in a magnetic field via mechanisms such as magneto resistance, which is a negative magneto-thermal conductivity, or the recently discovered thermal chiral anomaly which has a positive magneto-thermal conductivity(*1*). Magnons couple directly to magnetic fields and the thermal conductivity can be altered by the Zeeman effect red- or blue-shifting the magnon dispersion which alters the thermal occupation of magnon states. Phonons are generally considered not to be directly affected by a magnetic field, but we show here how magnon-phonon interactions can thoroughly change that picture.

Heat switches, also called thermal switches, are devices that alternate between good thermal conductors and good thermal insulators as needed. When one is installed in the heat-conduction pathway between a hot, heat-producing component and a heat sink, the change in thermal conductance can be used to control the component's temperature. The thermal conductance of the heat switch can be controlled by an external input such as electrical voltage, temperature, or external magnetic field. Solid-state heat switches that work without any moving parts are crucial in the realization of solid-state thermodynamic cycles such as spin-based heat engines and refrigerators. They are also an enabling technology for important efficiency improvements even in classical thermal-to-electricity conversion cycles, particularly those based on cyclical or intermittent sources of heat, such as sunlight or automotive exhaust(*2*). The switching ratio $SR = K_{\text{on}}/K_{\text{off}}$ is a figure of merit for a thermal switch, where $K_{\text{on}}$ is the highest thermal conductance and $K_{\text{off}}$ is the lowest thermal conductance that can be achieved by applying the control parameter. The lattice (or phonon) thermal conductivity is usually larger than the electronic and magnonic contributions. Therefore the lack of influence of magnetic field on the phonons makes it difficult to increase the switching ratio in all-solid-state heat switches unless mechanisms can be discovered that affect the lattice thermal conductivity by an external field.

Crystalline materials that integrate both magnetism and non-trivial topology of electronic band structure are of interest recently. One such material, $MnBi_2Te_4$, has attracted attention as a potential magnetic topological insulator. $MnBi_2Te_4$ crystalizes in a rhombohedral lattice of the space group $R\bar{3}m$ consisting of septuple van der Waals layers of Te-Bi-Te-Mn-Te-Bi-Te. This creates a structure that integrates a central layer of $MnTe_6$ octahedra inside the $Bi_2Te_3$ archetype, making it a magnetic relative of the 3D topological insulator $Bi_2Te_3$. $MnBi_2Te_4$ has an A-type antiferromagnetic (AFM) structure: the $Mn^{2+}$ spins have moments that are aligned in the out-of-plane direction, are ferromagnetically coupled within each layer, but are weakly antiferromagnetically coupled with neighboring layers. The Néel temperature is $T_N$ = 25 K. A-type AFM magnetic topological insulators such as $MnBi_2Te_4$ can contain both topological axion insulating states with AFM ordering and the quantum anomalous Hall effect, depending on the number of layers and whether an external magnetic field is applied. In an out-of-plane magnetic field with temperatures below $T_N$, the bulk magnetic ordering undergoes a spin-flop transition followed by a canted AFM (CAFM) ordering. Further increasing the field leads to a phase where the spins in all layers align, making $MnBi_2Te_4$ appear ferromagnetic (FM) at high field (depicted in Fig 1D) (*3*, *4*). The interplay between the magnetic structure and the topologically nontrivial bands endows the material with rich magnetic and topological phase transitions in field(*5–8*). $MnBi_2Te_4$ is therefore expected to exhibit nonmonotonic magneto-electrothermal transport phenomena, promising new mechanisms for magnetic field operated heat switches.

In this study we measure the in-plane thermal conductivity (κ) of the antiferromagnet $MnBi_2Te_4$ from 2 K to 30 K and find κ is modified by a magnetic field applied in the out-of-plane direction. Characteristic changes in the thermal conductivity coincide with the boundaries of the field-induced magnetic phase transitions in this material. κ decreases with field in the AFM phase, is approximately constant in the canted phase, and increases again with field in the FM phase. The magnitude and sign of the changes cannot be explained by either the magnon thermal conductivity or the electronic thermal conductivity. We propose instead that the magnon gap plays a crucial role in controlling the phase space of energy-momentum that allows magnon-phonon scattering(*9*). This understanding paves way for engineering functional materials for heat flow control. We also studied the potential for heat flow control in the transverse configuration. Thermal Hall data measured in the same configuration show an anomalous thermal Hall at the spin-flop transition which strongly resembles the electrical Hall data and relates to it via the Wiedemann-Franz law. Magneto-thermoelectric data are also reported for the first time in bulk $MnBi_2Te_4$.

## Results
### Thermal conductivity across phase transitions

Figure 1A shows the temperature dependence of the in-plane thermal conductivity $\kappa_{xx}$ without any applied magnetic field. This agrees with data previously reported in the literature (*10*). Upon cooling down from 200 K, $\kappa_{xx}(T)$ initially decreases and forms a shallow minimum at about 120 K. The electronic thermal conductivity was reported to follow a $T^1$ law(*10*), and can be estimated from the resistivity of the sample using the Wiedemann-Franz law with the free-electron Lorenz ratio to be of the order of 0.2 W m$^{-1}$ K$^{-1}$ at 100K. It is generally much smaller than the total $\kappa_{xx}$. At temperatures slightly above 30 K, which are above $T_N$ of 24.5 K, the lattice thermal conductivity dominates. Around $T_N$ and below, both the lattice thermal conductivity and the magnon thermal conductivity $\kappa_{lattice} + \kappa_{magnon}$ must be considered, while the electronic thermal conductivity diminishes and contributes less than 0.05 W m$^{-1}$ K$^{-1}$ to the total thermal conductivity of 2.62 W m$^{-1}$ K$^{-1}$. However, comparing the experimental data above and below the Neel temperature $T_N$ = 24.5 K, one notices an apparent suppression in $\kappa_{xx}$ below $T_N$ compared to what it would be if the data above $T_N$ were simply extrapolated following the $1/T$ law one expects for the Umklapp-dominated lattice thermal conductivity. Combined with the observation of a peak in heat capacity at the Neel temperature at zero field and absence of the peak at an applied magnetic field of 9 T(*11*), this suggests the emergence of magnons at $T<T_N$. An anomaly in $\kappa_{xx}(T)$ was observed in other magnetic materials near the ordering temperature (*12–14*). In yttrium iron garnet (YIG), the magnonic contribution to thermal conductivity was estimated to be up to ~1 W m$^{-1}$ K$^{-1}$ at 2 K and becomes a significant contribution as temperature decreases(*15*). If the magnonic contribution to thermal conductivity were significant in $MnBi_2Te_4$ in this range, we would see an increases of $\kappa_{xx}$ as the temperature decreases below $T_N$, yet we observe a decrease in this temperature range. We conclude that the magnons do not carry much additional heat but instead induce strong phonon-magnon scattering in the ordered phase.

The field dependence of the in-plane thermal conductivity $\kappa_{xx}(B_z)$ is shown in Figures 1B-D. Note that Figure 1B shows data in vicinity of $T_N$, while Figure 1C shows data at lower temperatures. As the temperature decreases towards the ordering temperature, $\kappa_{xx}(B_z)$ develops an interesting field dependence. At 29.2K, we observe a slight increase in $\kappa_{xx}(B_z)$ with an applied magnetic field up to 9 T. This magnetic field induced increase in thermal conductivity grows larger as the temperature approaches $T_N$. Below $T_N$, 22 K < T < 25 K, $\kappa_{xx}$ decreases at low field and increases linearly at high field. Far below $T_N$, T <

22K, the canted AFM ordering phase appears in the intermediate field region. Our data in Fig. 1C shows that $\kappa_{xx}$ saturates in this region with only a weak field dependence. In the FM phase, the field dependence becomes a linear increase with field. In Fig. 1D we plot the derivative $d\kappa_{xx}/dB_z$ as as function of $B_z$ and $T$ and overlay the known magnetic phase diagram (*3*, *4*). Discontinuities in $d\kappa_{xx}/dB_z$ coincide precisely with the magnetic phase boundaries, indicating that the changes in thermal conductivity are related to the magnetic phases. (*3*, *4*)

**Theory for the field dependence of thermal conductivity**

The field dependence of the thermal conductivity below $T_N$ is unlikely to be due to electrons, because they only contribute 0.05 W m$^{-1}$ K$^{-1}$ to $\kappa_{xx}$ and the resistivity data show less than a 2% change in a 9 T magnetic field at 25 K. Therefore, we must look at phononic and magnonic contributions. Above $T_N$, a magnetic field polarizes a paramagnet into a ferromagnetic state. Increasing the magnetic field stiffens the magnetic moments; thus, magnetic scattering is reduced. This explains well the $\kappa_{xx}(B_z)$ data above $T_N$.

The $\kappa_{xx}(B_z)$ trends below $T_N$ can be generalized as follows: A decrease in $\kappa_{xx}(B_z)$ in the AFM phase that becomes more linear at lower temperature, a sharp drop at the spin-flop transition, a lack of field dependence in the CAMF phase, and a linear increase with field in the FM phase. Strong suppression of thermal conductivity in the AFM phase and a sharp drop at the spin-flop transition were also reported in a multiferroic material(*16*) although the origin was not well established. A linear increase in thermal conductivity with magnetic field was also reported in Na$_2$Co$_2$TeO$_6$(*17*) and attributed to a reduction of magnon-phonon scattering. An increase in thermal conductivity at high field was observed in Bi-Sb topological insulators and attributed to the thermal chiral anomaly. This occurs when an applied magnetic field is colinear with the heat flux and parallel to the Weyl point separation(*1*). Although the FM phase of MnBi$_2$Te$_4$ is predicted to be a Type II Weyl semimetal with Weyl points separation from Γ-Z(*8*), in our experimental setup, the applied heat flux direction is perpendicular to the Weyl point separation, ruling out the thermal chiral anomaly. The theory of Fermi arc mediated entropy transport in Weyl semimetals(*18*) also predicts an increase of thermal conductance that is linear with an applied magnetic field that is perpendicular to the surfaces that host topologically protected Fermi arcs. In our experimental setup, it is possible that a small, unintentional misalignment of the out-of-plane magnetic field exists, so there may be a small in-plane magnetic field component $B_{\text{in-plane}}$ perpendicular to the arcs. However, no change was observed when the B$_{\text{in-plane}}$ component when we increased it intentionally by setting a small but intentionally misaligned angle between the applied magnetic field and the sample's out-of-plane direction (see Supplementary Materials), contrary to the theoretical prediction. In our samples, we note that the position of the Fermi level (see Methods section) of MnBi$_2$Te$_4$ is far (0.3 eV) from the bulk gap. Thus, the measured magneto-thermal transport behavior is unlikely to be due to topological properties. However, the distinct difference of the field dependence of $\kappa_{xx}$ in the different magnetic phases is striking.

To understand the behavior of the $\kappa_{xx}(B_z)$ data below $T_N$, we calculated the evolution of the magnon bands and inferred the consequences for magnon-phonon interactions. We used atomistic spin dynamics (see Supplementary Materials) based on the Heisenberg model parametrized from neutron diffraction measurements of MnBi$_2$Te$_4$(*19*). The Hamiltonian is

$$\mathcal{H} = -\frac{1}{2}\sum_{\langle ij\rangle_{\parallel}} J_{ij}\mathbf{S}_i \cdot \mathbf{S}_j - \frac{1}{2}\sum_{\langle ij\rangle_{\perp}} \mathbf{S}_i \cdot \mathbf{S}_j - \frac{1}{2}\sum_{\langle ij\rangle_{\perp}} J_c^{aniso} S_i^z S_j^z - D\sum_i (S_i^z)^2 - \sum_i \mu_s \mathbf{B} \cdot \mathbf{S}_i$$

where $i$ labels the Mn ions, $\mathbf{S}_i$ are unit vectors, $J_{ij}$ are the pairwise intralayer exchange interactions, $J_c$ is the nearest-neighbor interlayer exchange, $J_c^{aniso}$ is the nearest-neighbor interlayer anisotropic two-ion exchange, $D$ is the single-ion uniaxial anisotropy energy, $\mu_s = 5\mu_B$ is the size of the Mn magnetic moment in Bohr magnetons and $\mathbf{B}$ is the externally applied magnetic field in Tesla. The magnon-band dispersions in the ordered magnetic phases are calculated by solving the Landau-Lifshitz-Gilbert equation and calculating the spin-spin correlation functions in frequency and reciprocal space. The value of all parameters and the methods are detailed in the Supplementary Materials. Figure 2 shows the calculated magnon-band dispersions. The dashed lines qualitatively depict an acoustic branch of the phonon dispersion of MnBi$_2$Te$_4$ based on the monolayer phonon dispersion(6). Although the bulk phonon dispersion may deviate from this dispersion, the deviation is likely small, as both bulk MnTe and Bi$_2$Te$_3$ have similar acoustic phonon branches dispersing from 0 to 5 meV from the zone center to the edge. For reference, our experiment data was measured from 3 K to 25 K, corresponding to a $k_B T$ scale of 0.25-2.15 meV.

In zero magnetic field, the system is in the AFM phase and the magnons have a near linear dispersion (Fig. 2A). There are two modes with opposite polarization, but in zero field these are degenerate. A 0.6 meV gap at the zone center is induced by the magnetic anisotropy, enhanced by the exchange energy in antiferromagnets. An external magnetic field along $z$ breaks the symmetry between the spin-up and spin-down moments, thus lifting the degeneracy of the AFM magnon branches into two bands with a gap proportional to the external field strength (Fig. 2B-D). Increasing the magnetic field blueshifts one branch and redshifts the other. The redshifted branch becomes the dominant scatterer since it moves into a region with a higher magnon density of states. As the field increases, this branch becomes more important because the thermal occupancy of the high-energy branch decreases. Once the lower magnon mode has closed the energy gap at the zone center, further increasing the applied magnetic field causes an instability in the magnetic order, producing the spin-flop transition and the CAFM phase (Fig. 2E). In the CAFM phase, there is a gapless magnon branch attributed to a Goldstone mode and another high-energy branch. The gapless mode retains its dispersion throughout the CAFM regime without a dependence on the magnetic field. Above a critical field, the magnetic moments are forced to align with the magnetic field, and the FM phase is established with an acoustic and optical branch (Fig. 2F-H). The magnetic field increases the energy of both modes, opening a gap in the zone center proportional to the Zeeman energy, $g\mu_B B_z$.

Although the measured $\kappa_{xx}$ below 25 K has a contribution from magnon thermal conductivity, the trends of $\kappa_{xx}(B_z)$ can not be explained by magnon thermal conductivity based on the calculated magnon spectrum. The magnon gap closed by the field in the CAFM phase would result in a higher magnon density, thus increasing heat carrier density. Given the similar dispersion, magnon thermal conductivity would increase in the CAFM phase. This contrasts with the data presented in which $\kappa_{xx}$ approaches a minimum as the gap is closed, stays at the minimum when the gap is zero throughout CAFM phase and increases as the gap opens again.

To understand the changes in thermal conductivity in the different magnetic phases we consider the relationship between the magnon spectrum and the acoustic phonon dispersion. To first order, the dominant magnon-phonon interactions can be broken down into three classes: hybridization at crossing points of the dispersion, the magnon-number-conserving confluence process, and magnon-number-conserving Cherenkov scattering(*9*).

Magnon-phonon hybridization can occur at the touching points between magnon and phonon dispersions. The strength of magnon-phonon hybridization depends on the volume of the phase space at the touching point(*20*). In all cases here the bands simply cross and so the hybridization is likely to be weak. In the CAFM phase, there is no magnon gap for the lower branch; thus, hybridization cannot happen on the lower branch, yet the thermal conductivity forms a minimum in this regime, suggesting hybridization is not the dominant mechanism to explain the data. From the neutron scattering data(*19*), the hybridization was also not observed; again, suggesting that it is a weak effect.

Cherenkov scattering, which conserves the magnon number, can be expected in all magnetic phases. This process involves a magnon scattering into a phonon and a lower energy magnon. The scattering rate for this process depends on the detailed shape of the magnon dispersion but is allowed by energy and momentum conservation throughout the Brillouin zone. The scattering cross section will have some field dependence as the magnon dispersion changes with field, but no angular momentum is transferred to the lattice and both total energy and total linear momentum of the two magnons and one phonon are conserved; thus it cannot alter the thermal transport. Finally, we suggest that the magnon-number non-conserving confluence processes where both energy and angular momentum is transferred is the most relevant process in explaining our data. In this process, two magnons interact with a phonon. The process must obey the conservation of energy and angular momentum $\epsilon_{\mathbf{k}} + \epsilon_{\mathbf{k'}} - \omega_{\mathbf{q}\lambda} = 0$, where $\epsilon_{\mathbf{k}}$ is a magnon dispersion and $\omega_{\mathbf{q}\lambda}$ is the phonon dispersion with polarization $\lambda$. Crucially, this process is forbidden for phonons at energies less than twice the size of the magnon gap. These low energy phonons, corresponding to long wavelengths, typically can travel long distances across the lattice without scattering. They also have a very large thermal population according to Bose-Einstein statistics. Thus, they are the dominant heat-carrying phonons. As the magnon gap closes to zero in the CAFM phase, the confluence process is allowed everywhere in the Brillouin zone and interactions between magnons and the dominant heat-carrying phonons can occur. The zero gap in the CAFM phase also leads to a higher magnon density. In the end, two factors work together to explain the flat and lower thermal conductivity data in the CAFM phase: lower momentum phonons being scattered, and higher magnon density causing more scattering. These two factors have the opposite field dependence in the AFM and FM phases. In the AFM phase, the forbidden region becomes smaller with field, while in the FM phase, the forbidden region expands with field. This explains the opposite field dependence of thermal conductivity in these two phases.

**Thermoelectric and thermal Hall data**

Figure 3 shows the field dependence of the Seebeck, $S_{xx}(B_z)$, and Nernst, $S_{xy}(B_z)$, thermoelectric power at T < 30 K for the same sample. Both $S_{xx}(B_z)$ and $S_{xy}(B_z)$ are small in absolute value. The overall magnitude of $S_{xx}$ is on the order of a few μV K$^{-1}$ in the reported temperature range and a few tenths of μV K$^{-1}$ for $S_{xy}$. Although there are hopes for good thermoelectric properties of MnBi$_2$Te$_4$ due to its topological band structure, the small thermoelectric coefficients are consistent with a metallic system and are another result of the high Fermi level due to unintentional defect doping. Overall, the in-plane Seebeck and Nernst thermopowers in an out-of-plane magnetic field are consistent with the resistivity and Hall resistivity data (see Supplementary Materials). In the canted AFM phase below T$_N$, the Seebeck coefficient is slightly increased, in accordance with the slight decrease of resistivity and the Mott relation. The decrease in resistivity is attributed to the spin valve effect(*21*). The Nernst thermopower data show an anomalous Nernst effect and corresponds well with magnetization data. Below 20 K, $S_{xy}(B_z)$ has a small slope near zero field. At the spin-flop transition, $S_{xy}(B_z)$ jumps at the same point that the magnetization changes abruptly.

The thermal Hall effect, $\kappa_{xy}$, was also measured and is shown in Figure 4A. Above $T_N$, $\kappa_{xy}$ is a linear function of the field up to 9 T with the slope $d\kappa_{xy}/dB_z$ decreasing as the temperature decreases. Below $T_N$, $\kappa_{xy}$ shows an abrupt increase at the spin-flop transition. The thermal Hall conductivity resembles the electrical Hall resistivity (see Supplementary Materials). In Figure 4B, we show the thermal Hall conductivity $\kappa_{xy}$ calculated using the Wiedemann Franz law (WFL), $\kappa_{xy,WFL} = \sigma_{xy}L_0T$, where $L_0$ is the free electron value. The $\sigma_{xy}L_0T$ is about half of the measured value for $\kappa_{xy}$. Figures 4C and D show raw data points and averaged curves at T=10.6 K and 15.3 K, respectively. An increase in $\kappa_{xy}$ at 3.5 T is observed, coinciding with the spin-flop transition, and we attribute it to the anomalous thermal Hall effect. The close agreement within an order of magnitude indicates that the majority of thermal Hall conductivity is electronic in origin and is from the bulk. However, it is unexpected that the measured thermal Hall conductivity is larger than the estimated value using WFL, i.e., $\kappa_{xy} > \sigma_{xy}L_0T$. If the thermal Hall conductivity is purely electronic, it is shorted by lattice thermal conductivity, and we would expect $\kappa_{xy} < \sigma_{xy}L_0T$. The excess thermal Hall conductivity could be attributed to either a magnon contribution or a chiral phonon effect due to skewed magnetic scattering(*22*).

## Discussions

In summary, we show that MnBi$_2$Te$_4$ exhibits a significant and complex field-dependent magneto-thermal conductivity. We elucidate the mechanism to be one where phonons carry the heat and are subjected to magnetic scattering. Phonons carry most of the heat in most solids except the most metallic ones, are not generally thought of to be sensitive to magnetic fields. We showed that magnon/phonon interactions can induce an important field dependence to the amount of heat carried by the lattice, opening a new mechanism to realize heat switches, an enabling technology for solid-state heat engines and controlled cooling technologies.

While the topological band structure of MnBi$_2$Te$_4$ promises novel thermal transport properties such as a quantized anomalous thermal Hall effect, the Fermi level in these samples is far away from the band gap, and thus, it is unlikely that Weyl physics is involved in transport. Future studies on samples with greatly reduced doping levels may allow for the realization of Weyl-induced magneto-thermal transport phenomena.

## Materials and Methods

### Sample preparation

Single crystals of MnBi$_2$Te$_4$ were grown by adapting the previously established flux method (*10*) by slow cooling Bi$_2$Te$_3$ and MnTe powders in approximately a 5:1 ratio into an alumina Canfield crucible and centrifuging at 595 ºC. Crystals with lengths and widths of 3-8 mm and thicknesses of 10 – 200 µm were prepared for transport measurements.

### Sample characterization with Hall effect

Hall effect characterization of the carrier concentration (see Supplementary Materials) of the samples shows that electrons are the majority charge carriers. The electron concentration at 20K is about $6 \cdot 10^{19}$ - $1 \cdot 10^{20}$ cm$^{-3}$. This is very similar to other values reported, typically ranging from $7 \cdot 10^{19}$ - $1 \cdot 10^{20}$ cm$^{-3}$(*23*, *24*). The carrier concentration indicates that the Fermi level is about 0.3 eV into the conduction band(*5*). The n-type defects responsible for intrinsic electron doping were explored by Hou *et al.* and Du *et al.* both experimentally and computationally(*25*, *26*). Like $Bi_2Te_3$, $MnBi_2Te_4$ growth faces challenges with donor $Bi_{Mn}+$ anti-site defects (*23*, *27*), which heavily n-type dope the crystal.

**Experimental Design**

Thermal conductivity data was measured with the steady state method in the high-vacuum (10$^{-6}$ Torr), radiation-shielded chamber of a Quantum Design Physical Property Measurement System (PPMS). The heat source was a resistive heater (Omega Engineering, Inc., 120 Ω strain gauge) bonded to a copper heat spreader. The heat sink, also made of copper, was glued to the base of a PPMS's custom puck. The heat source and sink were bonded to the ends of the samples with silver epoxy. The thermometers were Cernox thermistors made by Lakeshore Cryotronics. The thermometers contacted the sample at different positions with epoxy. Resistance verses temperature calibration of the thermistors were done in-situ using AC lock-in method using Lakeshore 370 bridge and at different magnetic fields relevant to the study. Thermistor accuracy was tested to be better than 1 mK. We conducted measurements at discreet temperatures between 2 K and 300 K. The sample assembly was stabilized thermally at each discrete temperature for at least 30 minutes before measurement. Magneto-thermal conductivity was measured in magnetic fields from 9 T to -9 T in the PPMS. For angular-dependent magneto-thermal conductivity measurements, an angle between the temperature gradient and the magnetic field was created by bending the malleable heat sink at a desired angle. Magneto resistance and thermoelectric properties measurements were measured in separate measurements with additional current injection and voltage probing wires attached on to the sample.

**Error Analysis**

The thermal conductivity absolute measurement error is dominated by the sample geometry uncertainty, of the order of 10 %. Radiation heat losses were negligible in the temperature range of the reported magneto thermal conductivity data. Conduction heat losses via the thermistor wires were minimized by using long, thin (less than 40 μm diameter wires) manganin wires, which limited conduction heat losses through wires to 0.1% in the temperature range of magneto thermal conductivity measurements. Thermistor resistance was measured using low excitation to limit self-heating power on thermistors to be lower than 1 nW which was at least three orders of magnitude smaller than the heater power. (*28*, *29*). The relative errors, namely those in the temperature or field dependence, are much smaller and of the order of 2%: they are dominated by electrical noise in the thermometry and drift in the platform temperature.

**Acknowledgments**

We thank Dr. Brian Skinner for the helpful discussions. Simulations in this work were undertaken on ARC4, part of the High Performance Computing facilities at the University of Leeds, UK.

**Figures and Tables**

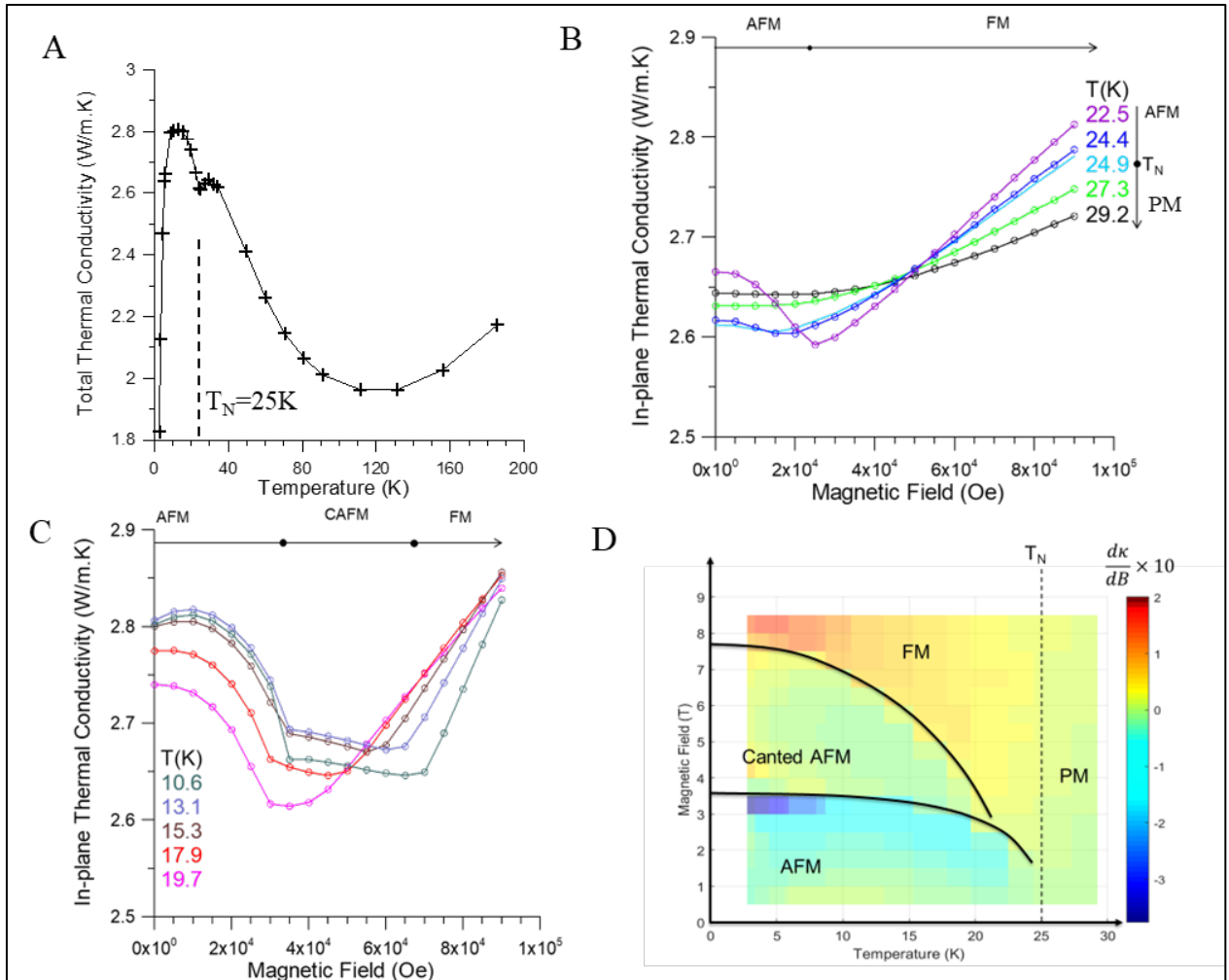

**Fig. 1. In-plane thermal conductivity and in-plane magneto thermal conductivity with out-of-plane magnetic field.** (A) Temperature dependence of total in-plane thermal conductivity $\kappa_{xx}$. $\kappa_{xx}$ decreases at the ordering temperature $T_N$=24.5 K indicating scattering of phonon to magnon. (B, C) Field dependence of in-plane thermal conductivity $\kappa_{xx}(B_z)$. Across the ordering temperature, $\kappa_{xx}(B_z)$ develops contrasting behavior at different field ranges. Above $T_N$, $\kappa_{xx}$ plateaus at low field then slightly increases with field. Below $T_N$, $\kappa_{xx}$ decreases with field at low field and increases linearly in field at high field. The magnetic field at which the field dependence changes correspond to the transition from AFM to FM ordering. Below 20 K, in addition to the initial decrease in AFM phase and linear increase in FM phase, there is a plateau in the intermediate Canted AFM ordering phase. (d) A magnetic ordering phase diagram can be reconstructed from a map of $d\kappa/dB$. The large negative value near the boundary of AFM phase corresponds to the spin-flop transition.

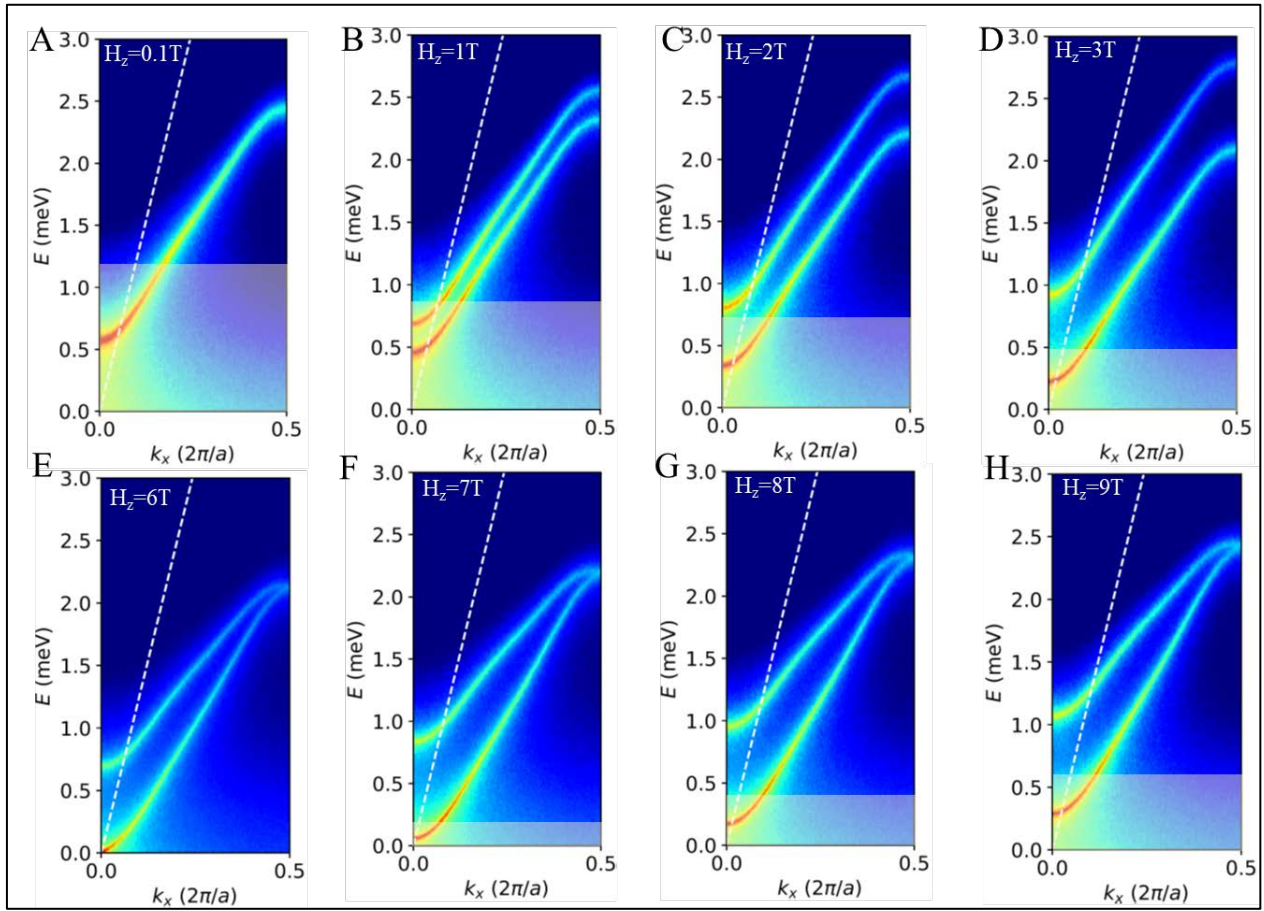

**Fig. 2. Calculated magnon band evolution in an out-of-plane magnetic field.** (A-D) magnon bands in the AFM ordering phase (E) magnon bands in the canted AFM ordering phase and (F-H) magnon bands in FM ordering phase. The dashed line qualitatively depicts an acoustic branch of phonon dispersion of $MnBi_2Te_4$ adopted from v14 phonon branch dispersion of $Bi_2Te_3$. The light-yellow boxes mark the forbidden region for magnon number non-conserving confluence process.

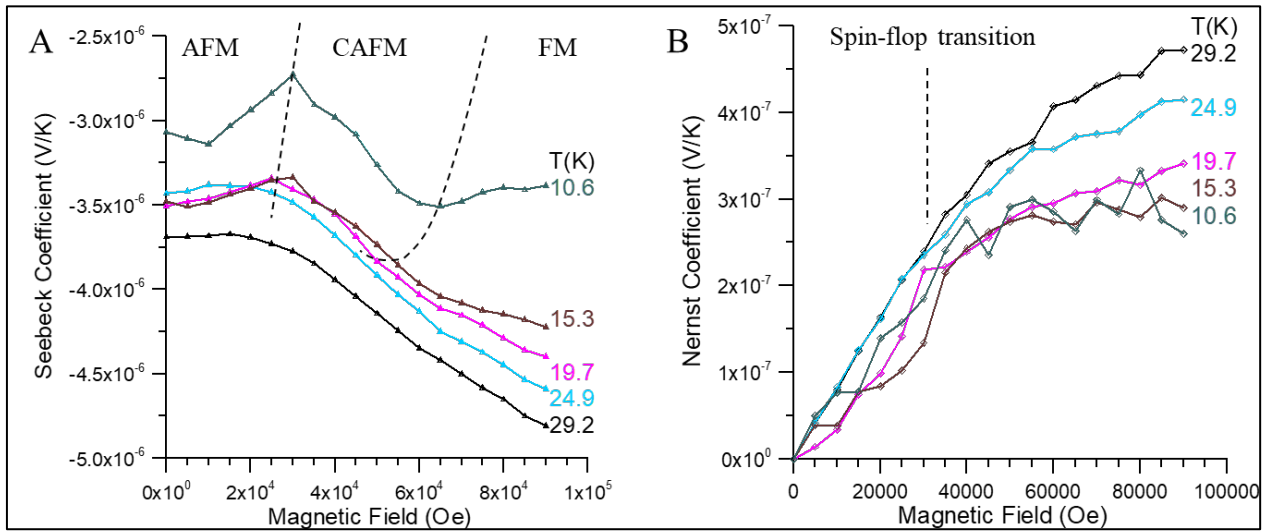

**Fig. 3. Field dependence of the Seebeck $S_{xx}(B_z)$ and Nernst $S_{xy}(B_z)$ thermoelectric power.** (A) Below $T_N$, the Seebeck coefficient shows an increase with $B_z$ in the CAFM phase and plateaus out at high field (FM). (b)The Nernst thermopower shows a sharp change of slope at the spin-flop transition below $T_N$. Comparing to the magnetization data, the Nernst data are characteristic of the anomalous Nernst effect.

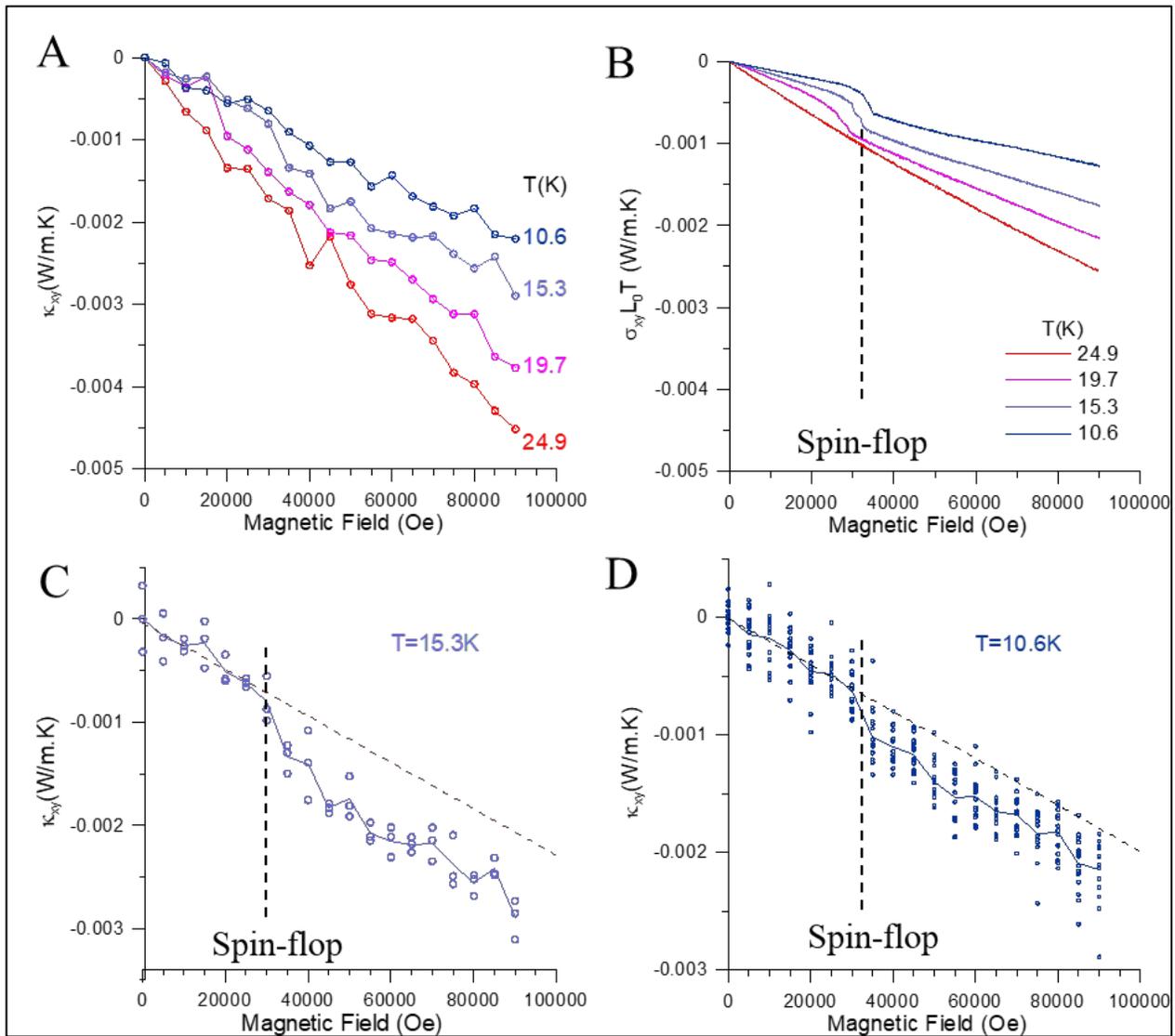

**Fig. 4. Thermal Hall conductivity and comparison with Wiedemann Franz Law.** (A) Thermal Hall conductivity $\kappa_{xy}$ verses applied magnetic fields measured below $T_N$. (B) Thermal Hall conductivity calculated from Wiedemann Franz Law $\kappa_{xy} = \sigma_{xy} L_0 T$ at corresponding temperatures. Quantitatively, measured thermal Hall conductivity is approximately twice as large as values predicted by Wiedemann Franz Law. (C, D) Blown-up plots of $\kappa_{xy}$ data at 15.3K and 10.6K shows an anomalous thermal Hall effect with a distinctive jump at the spin-flop transition. Individual data points are shown along with the lines connecting their mean values at each magnetic field. Dashed lines are linear fit drawn through the low field data points in the CAFM phase.

**Supplementary Materials**

Figs. S1. Resistivity and Hall resistivity data.

Figs. S2. Experimental test for presence of Fermi arc thermal transport.

Figs. S3. Magneto-thermal data at high temperature T≥50K.

Figs. S4. Magneto thermal conductivity data at T≤6K.

Figs. S5. Moment verses magnetic field on two samples.

Figs. S6. Thermal conductivity and magneto thermal conductivity of second sample.

Atomic Spin Dynamics

# Supplementary Materials for

## Magnetic field-controlled lattice thermal conductivity in MnBi$_2$Te$_4$

Dung Vu *et al.*

*Corresponding author. Email: heremans.1@osu.edu

**This PDF file includes:**

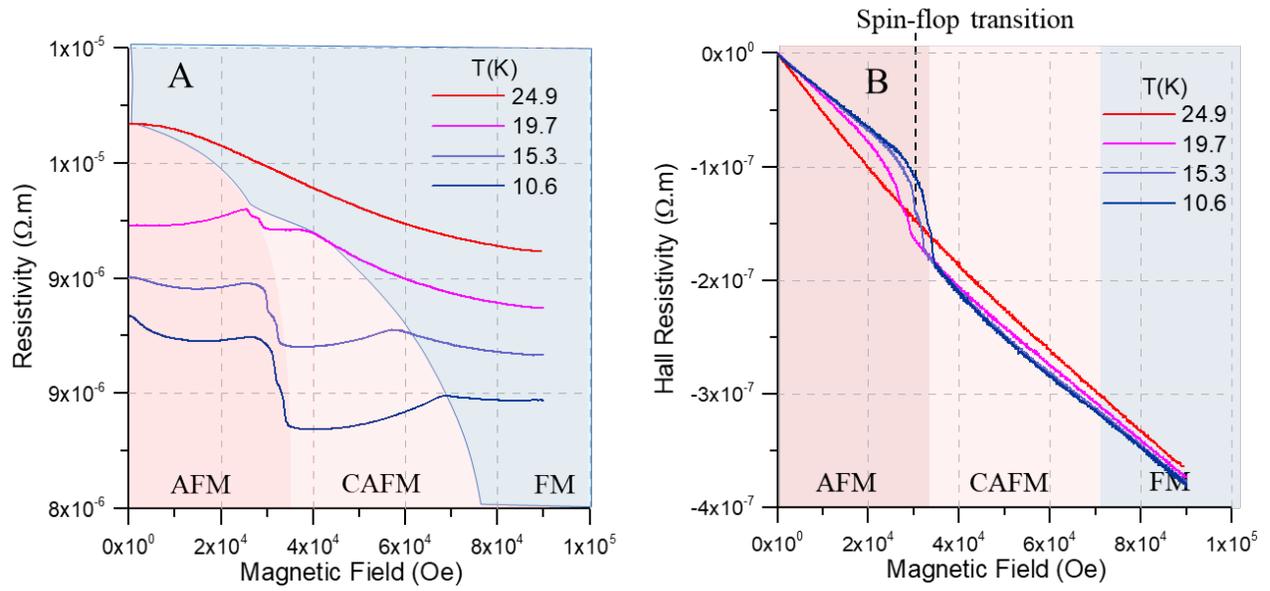

**Fig. S1.**

Resistivity and Hall resistivity data. (A) Field dependence of in-plane resistivity $\rho_{xx}(B_z)$. At Spin flop transition, the resistivity drop, attributed to spin valve effect. (B) Hall resistivity $\rho_{xy}(B_z)$ shows anomalous Hall effect at the spin-flop transition.

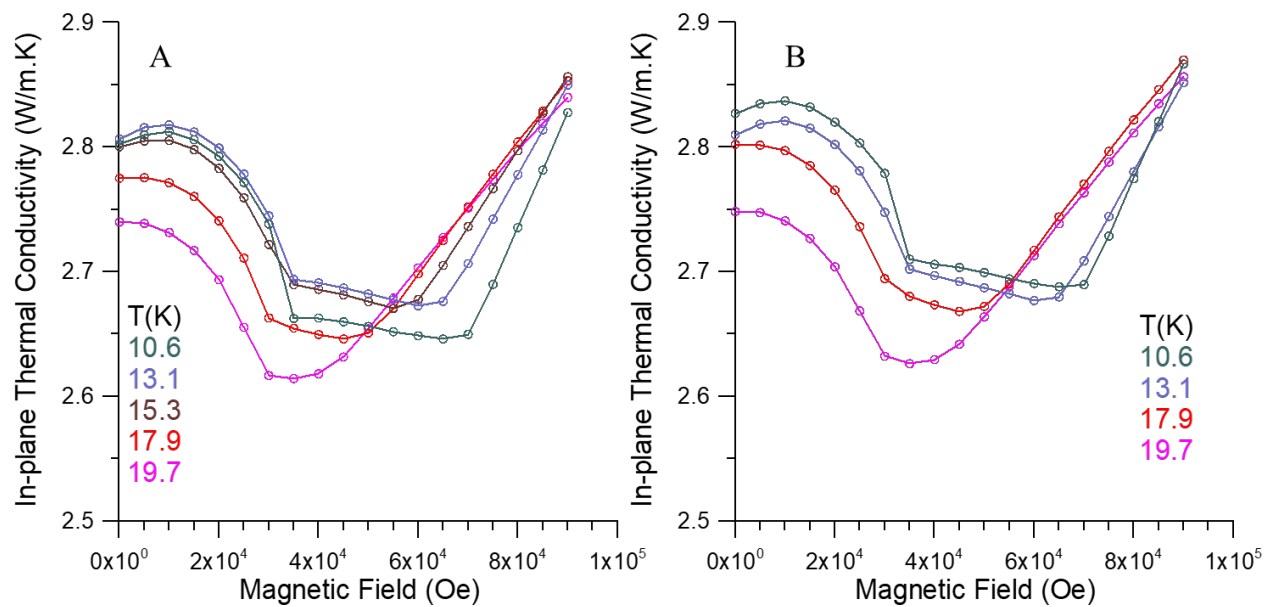

**Fig. S2.**

Experimental test for presence of Fermi arc thermal transport. (a) In-plane thermal conductivity with B in the out-of-plane direction (b) In-plane thermal conductivity with B 10° intentionally misaligned from the out-of-plane direction shows no change of slope in the FM phase.

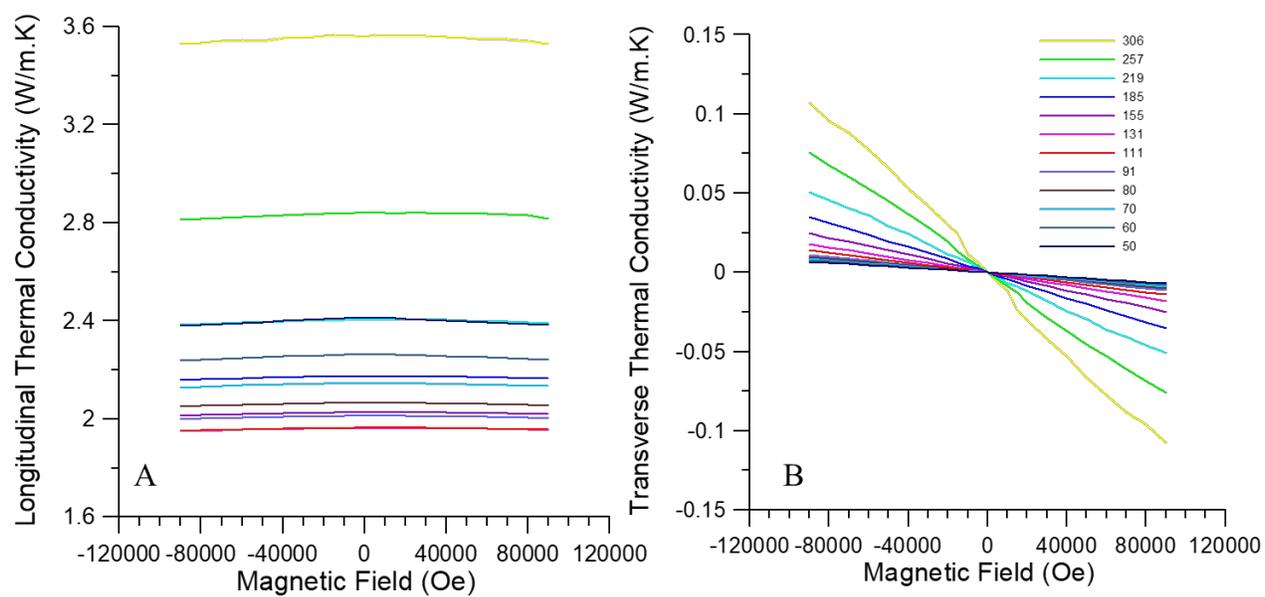

**Fig. S3.**

Magneto-thermal data at high temperature T≥50K. (A) Longitudinal magneto-thermal conductivity $\kappa_{xx}$ and (B) Thermal Hall conductivity $\kappa_{xy}$.

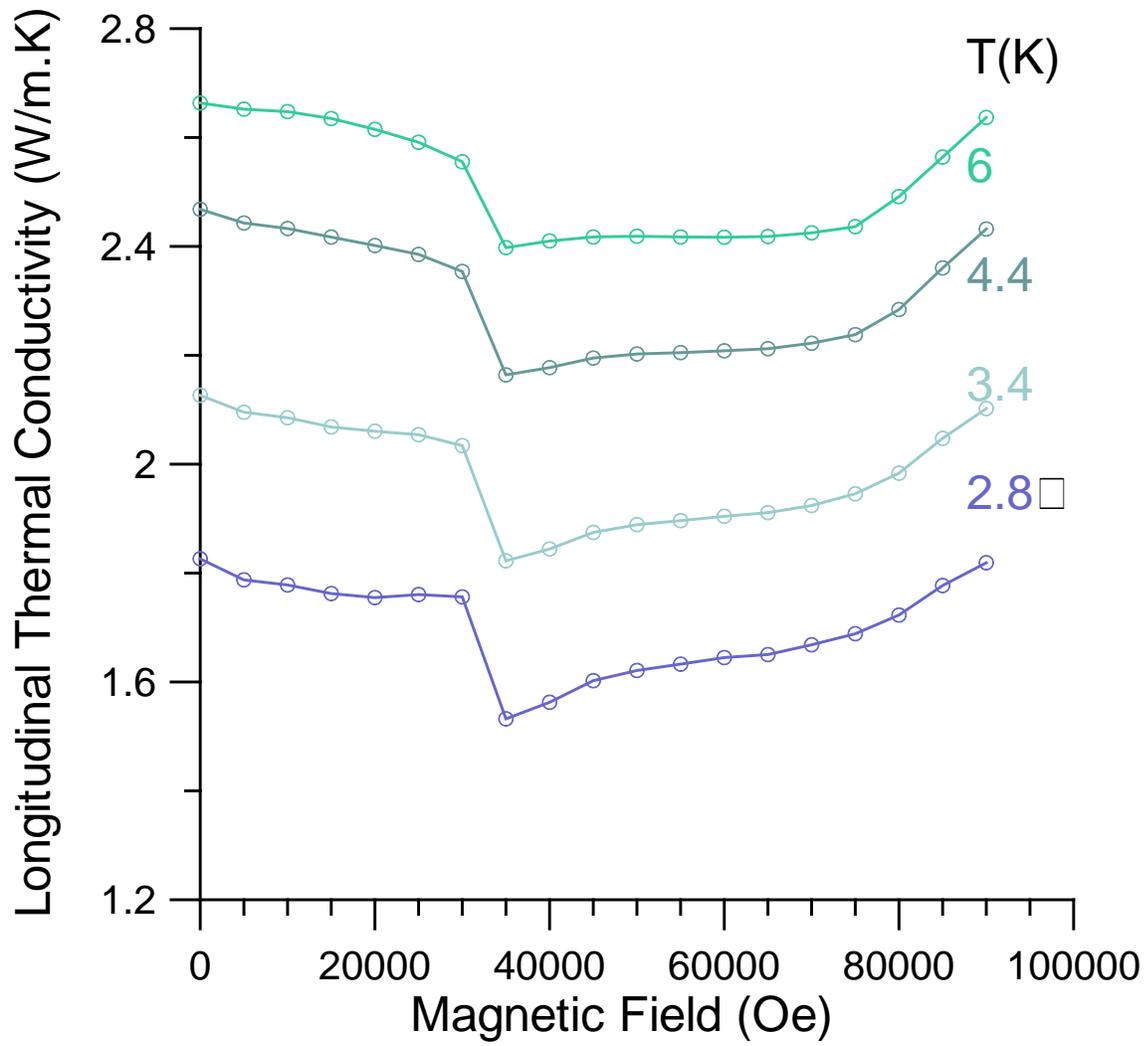

**Fig. S4.**
Magneto thermal conductivity data at T≤6K.

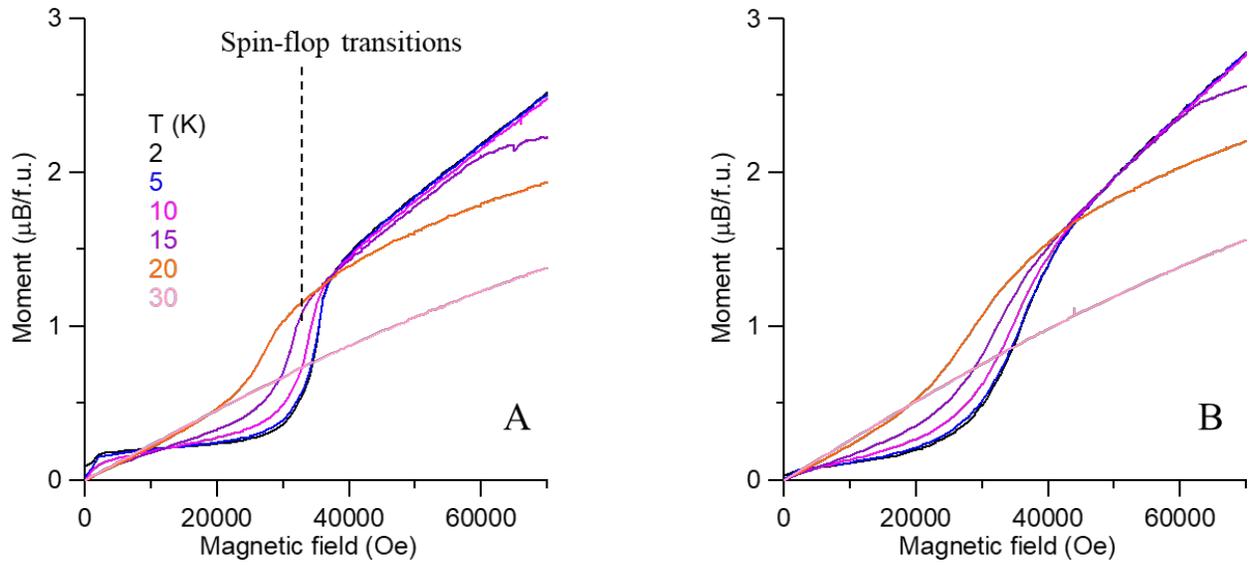

**Fig. S5.**
Moment verses magnetic field on two samples: (A) first sample with which data was reported in the main text shows sharp spin-flop transition in accordance with data reported in literature (B) second sample shows faint magnetic ordering transition and higher value of moment per formula unit, indicating the sample contains large number of free Mn atoms which act as paramagnetic free spins.

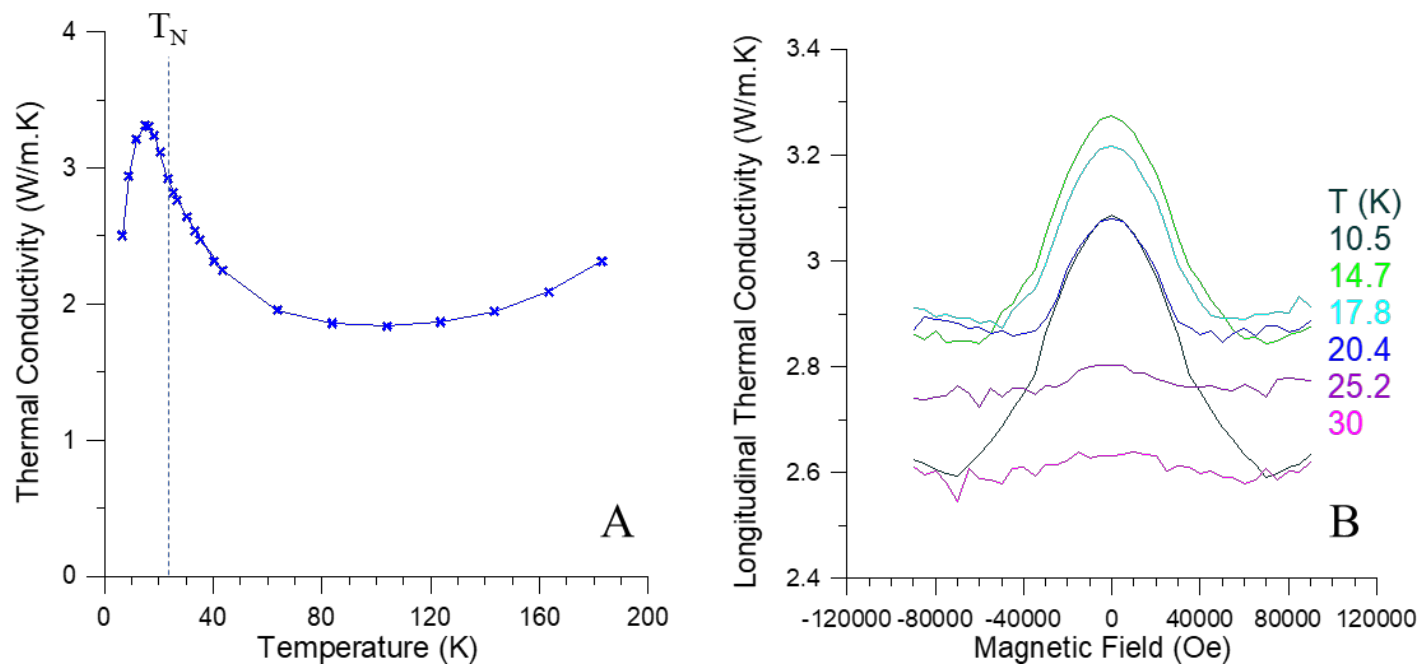

**Fig. S6.**
Thermal conductivity and magneto thermal conductivity of second sample. (A) Thermal conductivity of the second sample forms higher peak in at low temperature and no visible cusp at Neel temperature suggesting smaller magnon scattering and additional contribution from paramagnetic spins in this sample. (B) Magneto thermal conductivity of the second sample at temperatures from 10 to 30 K. While this sample still shows a faint change of dκ/dB at transitioning magnetic fields, these changes are small compared to the changes of dκ/dB in the first sample. This is because there is an additional large decaying thermal conductivity contribution from free spins that is constrained in an applied magnetic field.

# ATOMISTIC SPIN DYNAMICS

## Methods

MnBi$_2$Te$_4$ belongs to the R$\bar{3}$m (166) space group. The standard hexagonal unit cell is

$$\begin{aligned}\mathbf{a} &= (1,0,0) \\ \mathbf{b} &= (-1/2, \sqrt{3}/2, 0) \\ \mathbf{c} &= (0,0,18.88)\end{aligned} \quad (1)$$

where we have already doubled the **c** axis to account for the antiferromagnetic ordering (see Fig. S7) In the antiferromagnetic standard cell there are six Mn atoms, three of each of the antiferromagnetic sublattice (A and B). In fractional coordinates, these are located at

$$\begin{aligned}\text{Mn}_A &= (0,0,0) \\ \text{Mn}_B &= (2/3, 1/3, 1/6) \\ \text{Mn}_A &= (1/3, 2/3, 2/6) \\ \text{Mn}_B &= (0, 0, 1/2) \\ \text{Mn}_A &= (2/3, 1/3, 4/6) \\ \text{Mn}_B &= (1/3, 2/3, 5/6)\end{aligned} \quad (2)$$

We use the standard cell to compare with neutron scattering data and simulations in previous work by Li et al. (19) to validate the model parameters (see section ). However, for efficiency and clarity we performed the calculations in the main text using the primitive cell which has the basis vectors

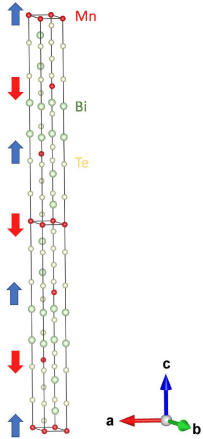

FIG. S7. Hexagonal unit cell of MnBi$_2$Te$_4$ with the antiferromagnetic ordering shown as red and blue arrows.

$$\begin{aligned}\mathbf{a} &= (\phantom{-}0.500000, \phantom{-}0.288675, \phantom{-}6.293360) \\ \mathbf{b} &= (-0.500000, \phantom{-}0.288675, \phantom{-}6.293360) \\ \mathbf{c} &= (\phantom{-}0.000000, -0.577350, \phantom{-}6.293360)\end{aligned} \quad (3)$$

In the primitive cell, there are only two Mn atoms, one of each of the antiferromagnetic sublattices. In fractional coordinates, these are located at

$$\begin{aligned}\text{Mn}_A &= (0.0, 0.0, 0.0) \\ \text{Mn}_B &= (0.5, 0.5, 0.5).\end{aligned} \quad (4)$$

We use the Hamiltonian and parameters for MnBi$_2$Te$_4$ suggested in Ref. 19 on the basis of linear spin wave theory fitting of inelastic neutron scattering measurements. The Hamiltonian written in our conventions is

$$\mathcal{H} = -\frac{1}{2}\sum_{\langle ij\rangle_\parallel} J_{ij} \mathbf{S}_i \cdot \mathbf{S}_j - \frac{1}{2}\sum_{\langle ij\rangle_\perp} J_c \mathbf{S}_i \cdot \mathbf{S}_j \\ - \frac{1}{2}\sum_{\langle ij\rangle_\perp} J_c^{\text{aniso}} S_i^z \cdot S_j^z - D\sum_i (S_i^z)^2 - \sum_i \mu_s \mathbf{B} \cdot \mathbf{S}_i \quad (5)$$

where factors of 1/2 are for the double counting in the sum and $\mathbf{S}_i$ are unit vectors. $J_{ij}$ are the isotropic intraplane exchange energies, $J_c$ is the isotropic interplane exchange energy, $J_c^{\text{aniso}}$ is the anisotropic interplane exchange energy, $D$ is a single-ion uniaxial anisotropy, $\mu_s$ is the size of the magnetic moment and **B** is an externally applied magnetic field in Tesla. $\langle \cdots \rangle$ indicates that a sum is performed only over a limited number of neighbours and $\parallel$ denotes inp-lane and $\perp$ out-of-plane neighbours. The parameters values and units are given in tables S1 and S2.

We solve the spin dynamics using the Landau-Lifshitz equation

$$\frac{d\mathbf{S}_i}{dt} = -\gamma\left[\mathbf{S}_i \times \mathbf{H}_i + \alpha \mathbf{S}_i \times (\mathbf{S}_i \times \mathbf{H}_i)\right], \quad (6)$$

where $\gamma$ is the gyromagnetic ratio and $\alpha$ is a damping parameter with the values given in Table S2. The effective field $\mathbf{H}_i$ on each lattice site is

$$\mathbf{H}_i = -\frac{1}{\mu_s}\frac{\partial \mathcal{H}}{\partial \mathbf{S}_i} + \boldsymbol{\xi}_i \quad (7)$$

with $\boldsymbol{\xi}_i$ being a vector of stochastic processes that provide thermal fluctuations. The temperature is introduced through a quantum thermostat that obeys the quantum fluctuation dissipation theorem (28) with the statistical properties of $\xi_i$ defined as

$$\langle \xi_{a,i}(t) \rangle = 0; \quad \langle \xi_{a,i}\xi_{b,j}\rangle_\omega = \delta_{ij}\delta_{ab}\frac{2\alpha}{\gamma\mu\beta}\frac{\hbar\omega}{e^{\beta\hbar\omega}-1}, \quad (8)$$

where $a$ and $b$ are Cartesian components, $\omega$ is the frequency, $\beta = (k_B T)^{-1}$ is the inverse thermal energy with $k_B$ is the Boltzmann constant and $T$ is the temperature, $\hbar$ is Dirac's constant, $\langle \cdots \rangle$ is a statistical time average and $\langle \cdots \rangle_\omega$ is a statistical average in frequency space. The stochastic process which produces this coloured noise is generated by solving a set of second-order stochastic partial differential equations (see refs. 28 and 29 for details). These equations and the Landau-Lifshitz equation are numerically integrated with a fourth-order Runge-Kutta integration scheme with a timestep $\Delta t = 1$ fs.

TABLE S1. Exchange constants for the Hamiltonian A5). $J_\parallel$ are intra-plane and $J_c$ are interplane interactions. A single interaction vector is given in fractional coordinates; equivalent vectors can be generated from the 3m space group operations. The interaction distances are given in units of the lattice constant $a$).

| symbol | vector (fractional coordinates) | number | distance (lattice constants) | value (meV) |
|---|---|---|---|---|
| $J_{\parallel 1}$ | ( 1, 0, −1) | 6 | 1.000000 | 0.5825 |
| $J_{\parallel 2}$ | ( 1, 1, −2) | 6 | 1.732051 | −0.0825 |
| $J_{\parallel 3}$ | ( 0, 2, −2) | 6 | 2.000000 | 0.0175 |
| $J_{\parallel 4}$ | ( 1, 2, −3) | 12 | 2.645751 | −0.0075 |
| $J_{\parallel 5}$ | ( 0, 0, 3) | 6 | 3.000000 | 0.0400 |
| $J_{\parallel 6}$ | ( 0, 2, 2) | 6 | 3.464102 | 0.0325 |
| $J_{\parallel 7}$ | ( 0, 1, 3) | 12 | 3.605551 | 0.0200 |
| $J_c$ | ( $\frac{1}{2}$, $\frac{1}{2}$, −$\frac{1}{2}$) | 6 | 3.199207 | −0.1625 |
| $J_c^{\text{aniso}}$ | ( $\frac{1}{2}$, $\frac{1}{2}$, −$\frac{1}{2}$) | 6 | 3.199207 | −0.0575 |

TABLE S2. Model parameters used in the Hamiltonian (5) and Landau-Lifshitz equation (6).

| Symbol | Value |
|---|---|
| $D$ | 0.2035 meV |
| $\mu_s$ | $5\mu_B$ |
| $\alpha$ | 0.05 |
| $\gamma$ | $1.76 \times 10^{11}\,\text{rad}\,\text{s}^{-1}\text{T}^{-1}$ |

**Validation of model parameters**

In Ref. 19 the spin vectors have length $S = 5/2$ but in our model the spin vectors are of unit length, and hence our exchange interactions are $J_{ij}^{\text{this work}} = SJ_{ij}^{\text{Li et al.}}$. The double counting convention and the exact definition of $\langle ij \rangle$ in Ref. 19 are unclear. Therefore, we calculated the neutron scattering cross section within our code (Fig. S8 and compared it with their experimental and simulated spectra to ensure that our models are equivalent.

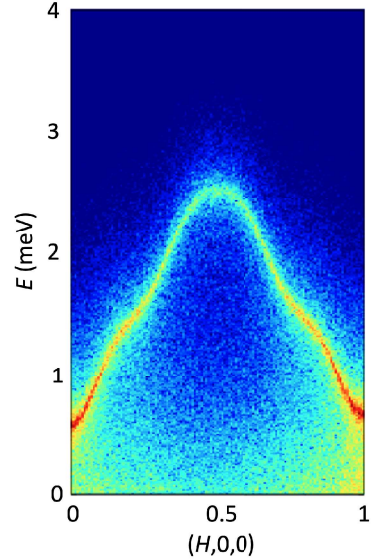

FIG. S8. MnBi$_2$Te$_4$ neutron scattering cross section at $T = 2$ K and $B_z = 0.1$ T to compare with Ref. 19 Fig. 2A and 2D.

Our neutron scattering cross section is calculated from

$$\mathcal{S}(\mathbf{Q},\omega) = \frac{g_n^2 r_c^2}{2\pi\hbar}\sum_{ab}\left(\delta_{ab} - \hat{Q}_a\hat{Q}_b\right)$$
$$\times \sum_{d,d'} e^{-i\mathbf{Q}\cdot(\mathbf{r}_d - \mathbf{r}_{d'})} \sum_{l,l'} e^{-i\mathbf{Q}\cdot(\mathbf{R}_l - \mathbf{R}_{l'})}$$
$$\times \int_{-\infty}^{\infty} e^{-i\omega t}\left[\langle S_{ld}^a(0) S_{l'd'}^b(t)\rangle - \langle S_{ld}^a\rangle\langle S_{l'd'}^b\rangle\right] dt, \quad (9)$$

where $a,b = \{x,y,z\}$ are Cartesian components, $\mathbf{R}_l$ is the position of the $l$-th unit cell, $\mathbf{r}_d$ is the $d$-th position in the unit cell, $g_n = 1.931$ is the neutron g-factor, $r_c = e^2/m_e c^2 = 2.8$ fm is the classical electron radius with $e$, $m_e$, and $c$ the elementary charge, the mass of the electron and the speed of light, $\mathbf{Q}$ is the scattering vector and $\hat{\mathbf{Q}} = \mathbf{Q}/|\mathbf{Q}|$. $\delta_{ab}$ is the Kronecker delta function. $\mathcal{S}(\mathbf{Q}, E) = \mathcal{S}(\mathbf{Q}, \hbar\omega)$.

**Magnon Spectrum Calculation**

To calculate the magnon spectrum containing all bands in the reduced zone scheme, we must avoid the phase cancellation caused by the magnetic structure factor and the projection onto a scattering vector. We first apply a rotation, $\mathbb{W}_d$, to each spin position $d$ of the unit cell:

$$\tilde{\mathbf{S}}_d = \mathbb{W}_d \cdot \mathbf{S}_d \tag{10}$$

such that the spin is rotated to align with the $z$-axis and the oscillations of the spin are about this axis. We then calculate the total dynamical structure factor as

$$\mathcal{W}_{ab}(\mathbf{k},\omega) = \frac{1}{2\pi} \sum_d e^{-i\mathbf{k}\cdot\mathbf{r}_d} \sum_{l,l'} e^{-i\mathbf{k}\cdot(\mathbf{R}_l - \mathbf{R}_{l'})}$$
$$\times \int_{-\infty}^{\infty} e^{-i\omega t} \left[ \left\langle \tilde{S}^a_{ld}(0) \tilde{S}^b_{l'd'}(t) \right\rangle - \left\langle \tilde{S}^a_{ld} \right\rangle \left\langle \tilde{S}^b_{l'd'} \right\rangle \right] dt. \tag{11}$$

where $\mathbf{k}$ is a reciprocal space vector in the first Brillouin zone. To produce the figures in the main text, we plot the transverse components of the dynamical structure factor $(\mathcal{W}_{xx}(\mathbf{k},\omega) + \mathcal{W}_{yy}(\mathbf{k},\omega))^{1/2}$.

---